# Online Monitoring of Global Attitudes Towards Wildlife

Joss Wright, Robert Lennox, Diogo Veríssimo

## Abstract

Human factors are increasingly recognised as central to conservation of biodiversity. Despite this, there are no existing systematic efforts to monitor global trends in perceptions of wildlife. With traditional news reporting now largely online, the internet presents a powerful means to monitor global attitudes towards species. In this work we develop a method using the Global Database of Events, Language, and Tone (GDELT) to scan global news media, allowing us to identify and download conservation-related articles. Applying supervised machine learning techniques, we filter irrelevant articles to create a continually updated global dataset of news coverage for seven target taxa: lion, tiger, saiga, rhinoceros, pangolins, elephants, and orchids, and observe that over two-thirds of articles matching a simple keyword search were irrelevant. We examine coverage of each taxa in different regions, and find that elephants, rhinos, tigers, and lions receive the most coverage, with daily peaks of around 200 articles. Mean sentiment was positive for all taxa, except saiga for which it was neutral. Coverage was broadly distributed, with articles from 73 countries across all continents. Elephants and tigers received coverage in the most countries overall, whilst orchids and saiga were mentioned in the smallest number of countries. We further find that sentiment towards charismatic megafauna is most positive in non-range countries, with the opposite being true for pangolins and orchids. Despite promising results, there remain substantial obstacles to achieving globally representative results. Disparities in internet access between low and high income countries and users is a major source of bias, with the need to focus on a diversity of data sources and languages, presenting sizable technical challenges. Tackling these will depend on the development of more advanced natural language tools in a wildlife conservation context, as well as the development of systems to access more, and more diverse, data sources.

# Introduction

Biodiversity conservation requires detailed understanding of the population trends of species across time and space (Marsh & Trenham 2008). This type of information is central to outputs such as the IUCN Red List or the WWF Living Planet Index, which have been key to shaping our understanding of the biodiversity crisis and the policies that try to address it ("Indicators for the Strategic Plan for Biodiversity 2011-2020 and the Aichi Biodiversity Targets" n.d.). Because of this recognition, the last decades have experienced an important investment in the long-term monitoring of changes in species presence, abundance, and density across hundreds of sites worldwide. Iconic examples of this effort are the Audubon Christmas Bird Count, which has been ongoing for 120 years. A newer example is the iNaturalist App that has generated more than 25 million species presence records.

There is now a growing understanding that human factors are central to the effective conservation of wildlife (Byerly et al. 2018) . Yet, there has not been a comparable effort around the monitoring of human aspects such as attitudes and social norms either in extended temporal or large spatial scales, making it hard to gather key insights that can help in the management and conservation of wildlife. There are multiple likely reasons for this. In temporal terms, the study of human perceptions of wildlife is made intrinsically complex as it brings together not only the enormous diversity of life on Earth but also the vast heterogeneity amongst human groups and the rapid pace at which societies change over time (Renfrew 1994; Butler 2016). Measuring societal perceptions and interest in species requires extensive and costly social surveys that typically have limited geographic coverage (Fienberg & Tanur 1983). Despite being an essential component of biological conservation, few long-term and large-scale datasets of human attitudes and social norms exist concerning wildlife. The existing datasets that fulfil these conditions are either in the private domain, such as those collected by private polling companies, or suffer from a focus on a very narrow set of geographies and species (Treves et al. 2013), which make them less strategic for the understanding of the complex geopolitical relationships between human stakeholders and wildlife at national and global scales.

Advances in both the development and the adoption of new information technology creates an unprecedented opportunity to collect insights about human relationships with wildlife at scale. First, the expansion of the internet, with 2019 marking the first time more than half of the world's population was online (International Telecommunications Union 2019). This expansion means that not only are there an increasing number of users generating data via platforms such as social networks, but also that even traditional media are increasingly online. These sources of public data, coupled with techniques such as sentiment analysis, offer a wealth of opportunities when it comes to understanding attitudes and social norms surrounding wildlife.

These trends have not gone unnoticed to marketers who have increasingly worked to develop tools to harness public opinion data generated online and better understand consumers. While conservationists have been slow to adapt these techniques (Fidino et al. 2018), the field of conservation culturomics (Ladle et al. 2016), which seeks to understand human culture through the quantitative analysis of changes in word frequencies in large bodies of digital texts, has recently been highlighted as a key emerging area of conservation science (Sutherland et al. 2018).

One key area of methodological development has been sentiment analysis, which consists of automatically extracting and quantifying affective states from large bodies of text (Liu 2012). In this way, it is possible to extract not only information about the polarity of the text, whether it is positive or negative, and emotional states such as angry, sad, or happy expressed in the text; but also to link those affective measures to specific topics and the people or entities expressing those opinions. When applied to large bodies of text produced by the millions of users across social media platforms as well as online media repositories covering the most influential newspapers and news agencies, sentiment analysis can be used to monitor changes in attitudes and social norms towards species simultaneously at national, regional, and global scales. Along with geolocation of the information sources, this approach can produce spatially and temporally explicit data, allowing the characterisation of both temporal and spatial trends.

We explore how sentiment analysis can be used to understand both digital media and user generated data to obtain real time measurements of the attitudes towards different species of wildlife at multiple spatial scales. This would level the human dimensions of wildlife with the ecological side of wildlife conservation by making it possible to understand trends over time.

In this work, we describe and deploy a system to monitor the media saliency and attitudes towards wildlife using as a case study of seven taxa that have been involved in the highly salient issue of the illegal wildlife trade. We then discuss some of key limitations of this approach, related both to coverage and representation of different socioeconomic groups induced by the use of online data sources, and specific platforms; as well as the caveats of existing sentiment analysis tools and online datasets, namely around linguistic diversity and specific use of terms when discussing wildlife. Finally, we identify ways forward and highlight the opportunities that massive, online, realtime data sources present for biodiversity conservation.

## Monitoring online attitudes towards wildlife

*Data*

The core data source used in this analysis is the Global Database of Events, Language, and Tone (GDELT n.d.), which indexes over 189,000 global broadcast, print, and online news media from 241 countries or territories. GDELT includes news articles in over 100 different languages, of which 65 are machine-translated into English for processing, allowing for text-based search of articles in multiple languages based on English language keywords.

GDELT offers a number of services, many of which are situation-focused, tracking emerging events such as conflict or natural disasters. For our purposes, however, the key capability offered by GDELT is its full-text search of a vast corpus of online news media. The dataset GDELT's database of online news articles is continually updated, with full-text search available for articles released within the previous three days. GDELT does not provide access directly to the full text of articles, but instead provides links to the online article source, requiring the original text to be retrieved.

There are a number of challenges in searching, retrieving, and processing such large volumes of unstructured data. We discuss here four key steps of the process: search, retrieval, filtering, and processing.

*Search*

Appropriate selection and definition of search terms is crucial to achieving coverage of a species of interest whilst avoiding unacceptably high levels of false positive results. An online news search for *bear*, for example, may well return results for US sports teams *('Chicago Bears')*, television personalities *('Bear Grylls')*, and even financial news *('bear markets')*.

In many online searches this problem can be alleviated through the use of *negative keywords*: in most online search engines a query for '*bear -football -chicago*', for example, would largely remove results related to the Chicago Bears. Unfortunately, full-text keyword search in GDELT, and in a number of other large-scale online platforms such as Twitter, does not allow for such negative keywords in search.

To counter this problem we define *families* of search terms, one for each species. Each family contains a series of search terms that include both the species keyword and a qualifying keyword such as *conservation* to reduce the chance of matching unwanted articles (gathering news on different taxa from the Global Database of Events, Language, and Tone (GDELT; www.gdeltproject.org)).

*Table 1: Search terms used in gathering news on different taxa from the Global Database of Events, Language, and Tone (GDELT; www.gdeltproject.org)*

| Main Keyword | Additional Keywords |
| --- | --- |
| elephant | ivory, poach, wildlife, conservation, animal, seized, seizure, asian, african |
| rhino | horn, poach, wildlife, conservation, animal, black, white, seizure, seized |
| pangolin | scale, poach, wildlife, conservation, animal, anteater, seizure, seized |
| saiga | horn, seizure, seized, poach, wildlife, conservation, animal, antelope |
| tiger | bone, skin, seizure, seized, poach, wildlife, conservation, animal, bengal, cat |
| lion | bone, skin, seizure, seized, poach, wildlife, conservation, animal, cat |
| orchid | flower, ornamental, collector, wildlife, conservation, plant, flower, phalaenopsis, seized, seizure |

In each case, results returned from multiple searches are deduplicated to avoid both unnecessary download of the full data, and also to prevent double-counting of articles for individual groups of search terms. Our search tool makes use of the *gdeltr2* R package (Bresler n.d.), which allows direct querying of the GDELT data from the R programming language.

*Retrieval*

The search process above produces a list of articles matching any one of the search terms included in a search family. The data structure returned by GDELT includes a substantial number of features, of which we store only a subset relevant to our analysis. The most important details for our purposes are the date of article publication, the country of publication, the language of the article, and the URL at which the article was published. Notably, GDELT does not store or present the text of the articles themselves.

Following retrieval of the search results, our system follows the URL of the article and extracts the raw text of the page at which it is presented. If available, GDELT returns URLs for articles formatted both for full web browsers and for access on mobile devices such as smartphones, where available. We favour mobile-formatted URLs by default to download, as the presentation of these is typically simplified to meet the requirements of smaller screens, resulting in less extraneous text and interactive elements on the page. If, however, a mobile URL is not available, we download the desktop version of the article page. All web scraping takes place via the *rvest* web scraping framework for R (Wickham & RStudio 2019). Having retrieved the article for each returned

URL, we remove all formatting and store the raw text of the article in a database alongside the date, search terms, country, and language of the article.

*Filtering*

Despite careful selection of keywords to avoid overly-inclusive search terms, the nature of simple keyword-based searches inevitably returns a proportion of irrelevant articles, which can be high where keywords overlap significantly with other topics. To counteract this, and avoid polluting the database with a high number of extraneous articles, we apply a filtering procedure to the database after article retrieval. The tool we use is a *naïve Bayesian classifier* (Hastie et al. 2009), a well-known machine learning approach commonly used in email spam filtering systems.

To apply the naïve Bayesian filter, we must first train the algorithm to classify articles according to their relevance to conservation. To do so, we extracted a subset of articles gathered by our media scraping system and manually tagged each as either relevant or irrelevant to our analysis. This manually classified *training set* of articles is input into the filter, from which the system learns those textual features that distinguish relevant from irrelevant content. Following training we can asssess the accuracy of the filter by applying it to a second manually tagged set of articles and calculating the extent to which the trained filter agrees with the human-classified data.

Accuracy in machine learning tasks such as classification is typically expressed in terms of two quantities: *precision* and *recall* (Rijsbergen 2004). *Precision* refers to the proportion of documents correctly classified by the machine as relevant, excluding those that are irrelevant. *Recall*, by contrast, refers to the number of documents from the total corpus that are successfully classified. Both scores are needed to assess the effectiveness of a classification approach, as each is susceptible to a particular type of error: a classification approach with high precision will be unlikely to include irrelevant documents, but may create false negatives by falsely removing some relevant documents. By contrast, an approach with high recall will be sure to identify almost all relevant documents, but may also falsely include irrelevant documents. Ideally, both precision and recall would be high for a given approach, but a specific application may be more open to one or the other being lower.

$$precision = \frac{|\{relevant\ articles\} \cap \{all\ articles\}|}{|\{all\ articles\}|}$$

$$precision = \frac{|\{relevant\ articles\} \cap \{all\ articles\}|}{|\{relevant\ articles\}|}$$

For the purposes of this analysis we manually coded a training set of 1,200 articles according to their relevance to conservation topics. The accuracy of the trained filter was then validated on a further test set of 400 manually coded articles. The results demonstrated an acceptable balance between removing a high proportion of irrelevant articles without an undue number of false negatives that mark relevant articles as irrelevant.

Our naïve Bayesian filter achieved a precision of 0.765 and a recall of 0.962 when applied to previously unseen articles. This score provides us with a relatively high confidence that few relevant articles were rejected by our filter, although with some possibility of irrelevant articles being included in the final results.

The benefits of the Bayesian filtering step are significant: of 30,892 articles retrieved in the nine month period from the 6th May 2019 to the 28th January 2020 that matched our set of *elephant* search terms, only 9,527 were classified as relevant. Without application of the filter, approximately two-thirds of the results returned by a simple keyword search are not conservation-related, and would consequently have significantly biased our overall analysis. For such wide-scale analysis of media data, therefore, this filtering step is crucial.

***Processing***

With articles retrieved from the GDELT database and filtered to reduce irrelevant content, we applied sentiment analysis to the resulting text. For this purpose we make use of the *sentimentr* package (Rinker 2019).

There are a variety of approaches to analysing the sentiment of texts. The most common approach, however, and the one applied here makes use of standardised lexicons in which each word is scored, based on manual analysis, with an overall sentiment marking it as a positive or negative term. Words such as *happy* or *good* result in positive sentiment, whilst terms such as *sad* or *bad* result in negative sentiment. The overall sentiment score for a text is calculated by a weighted average of key terms in the document. More advanced sentiment analysis packages, such as *sentimentr*, also account for *valence-shifters* such as *not* that can reverse the sentiment of a word allowing, for example, "not good" to be correctly recognized as a negative sentiment. The resulting sentiment score for a given article is not only positive or negative, but also reflects the *intensity* of the sentiment, allowing an extremely positive text to be distinguished from one which is overall positive, but largely neutral.

This lexicon-based approach to sentiment analysis, and indeed almost all forms of sentiment analysis, require a training set of words or texts that have been manually tagged with sentiment scores. As this is an extremely labour-intensive process, standardised sentiment analysis dictionaries are used in the majority of tasks. Due to the small number of comparable sentiment analysis tools for other languages, and the lack of accuracy of automated machine translation across the range of languages covered in GDELT, we restrict ourselves to English-language news articles in this work whilst noting the need for much broader linguistic coverage.

***Temporal trends***

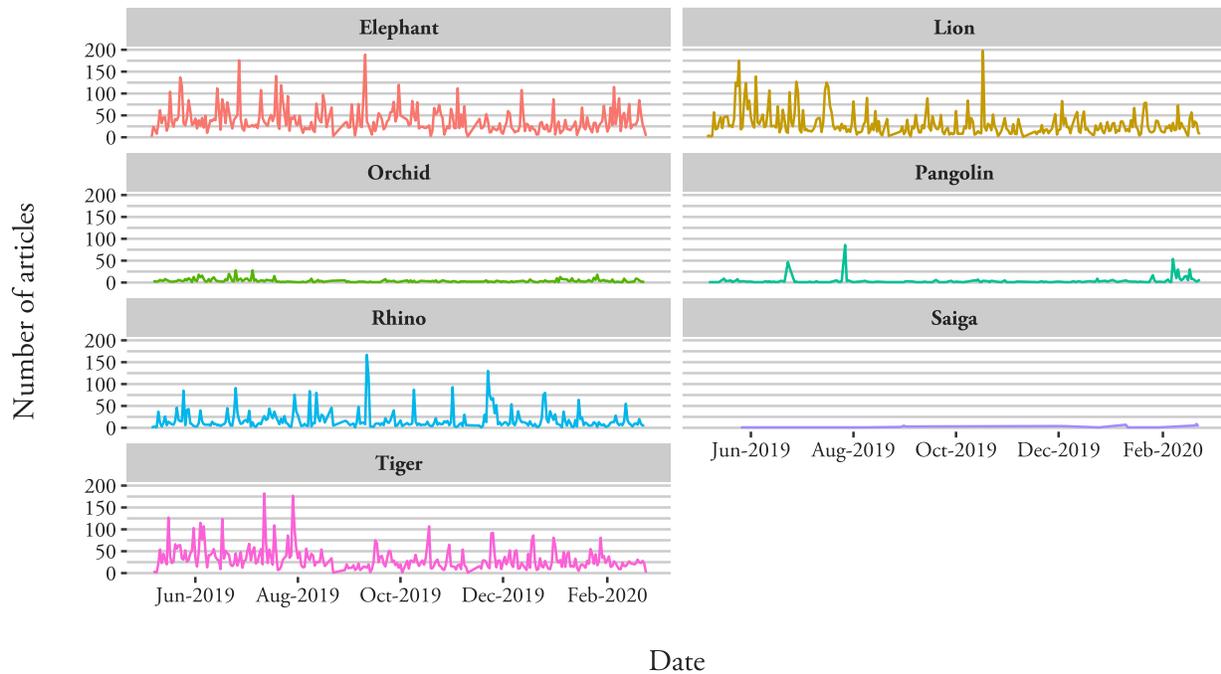

*Figure 1: Article count per day for target taxa based on the Global Database of Events, Language, and Tone (GDELT; [www.gdeltproject.org](www.gdeltproject.org))*

The number of times a subject is referred to in the media, known as *media saliency*, has been used in political science and more recently in the environmental context as a proxy for the extent to which that topic is found to be important by the public (McCombs & Shaw 1993; Veríssimo et al. 2014). In our data set, the overall media coverage varied greatly between days and amongst taxa. All taxa had days on which they received no coverage, although all taxa except saiga received some coverage in the majority of days. Orchids received coverage on 80% of the days considered in this study; lion, rhino, elephant, and tiger were covered on more than 90% of studied days. The taxa with the greatest number of articles published both overall and in a single day were tiger, elephant, and lion, which also had the highest mean number of articles published per day, at 31, 36, and 29 respectively.

It was unsurprising to see large felids and pachyderms receive the most media coverage. This is likely not only related to their longstanding status as some of the most charismatic animals on Earth (Veríssimo et al. 2017) but also due to the controversies these taxa have been involved in when it comes to issues such as wildlife trafficking, hunting, and human-wildlife conflict. Orchids, however, received only marginal media coverage despite being a group with more than 20,000 species that is heavily traded globally (Hinsley et al. 2018). This is, however, not surprising given the well documented lack of importance that plants are afforded around the topic of wildlife trade and more broadly in biodiversity conservation (Margulies et al. 2019). Another taxon that received little attention was the saiga, a species that, while traded internationally, most frequently receives coverage due to mass die-offs events (Nicholls n.d.). The coverage focused on pangolin was substantially lower than that from more traditionally charismatic taxa, which is surprising given how much the taxa has been heralded as the flagship species for the illegal wildlife trade (Harrington et al. 2018). The coverage around pangolins did seem to increase in 2020, with about half the articles being published during that time period, despite it only accounting for about40% of the articles overall and less than 20% of the time. It remains to be seen if this pattern will be retained in the long term.

Sentiment was narrowly positive across all taxa, with mean sentiment across all articles varying between 0.03 for pangolin, to 0.1 for orchids. The exception was saiga, for which sentiment was neutral, with a mean of 0 across all articles. This result was surprising given that most sources in GDELT are from North America and Europe, where we would expect the press to cover animals such as elephants or tigers positively, given their citizens do not have to bear the costs of sharing landscapes with these taxa. That said, the overall sentiment in media reporting may have been affected by stories around poaching or trafficking, which are topics that are widely covered in those regions and that tend to have a more negative sentiment.

All taxa ranged between positive and negative daily sentiment values, with the pangolin having the lowest daily mean sentiment in mid-June 2019 (-0.3) and orchids having the highest daily sentiment (0.45), recorded in mid-September 2019. Pangolins saw the largest variation in daily sentiment, with scores between -0.3 and 0.38, while elephants had the narrowest variation in daily sentiment, with scores varying between -0.18 and 0.18.

These results represent the range of attitudes associated with these particular taxa. Pangolins likely see negative sentiment due to their relative obscurity beyond their alleged label of the most trafficked animal in the world. This, in turn, focuses news reporting on massive seizures of both scales and whole animals, commonly dead, which drives a negative sentiment in the reporting. Orchids, whilst being one of the most trafficked taxa in the world, seem to be most commonly featured in the news due to their aesthetics or interesting ecological features, resulting in positive sentiment on average. This feature, in particular, highlights that sentiment of reporting may not represent perceptions of a given taxa itself in a positive or negative light, but instead the topics associated with reporting of that taxa.

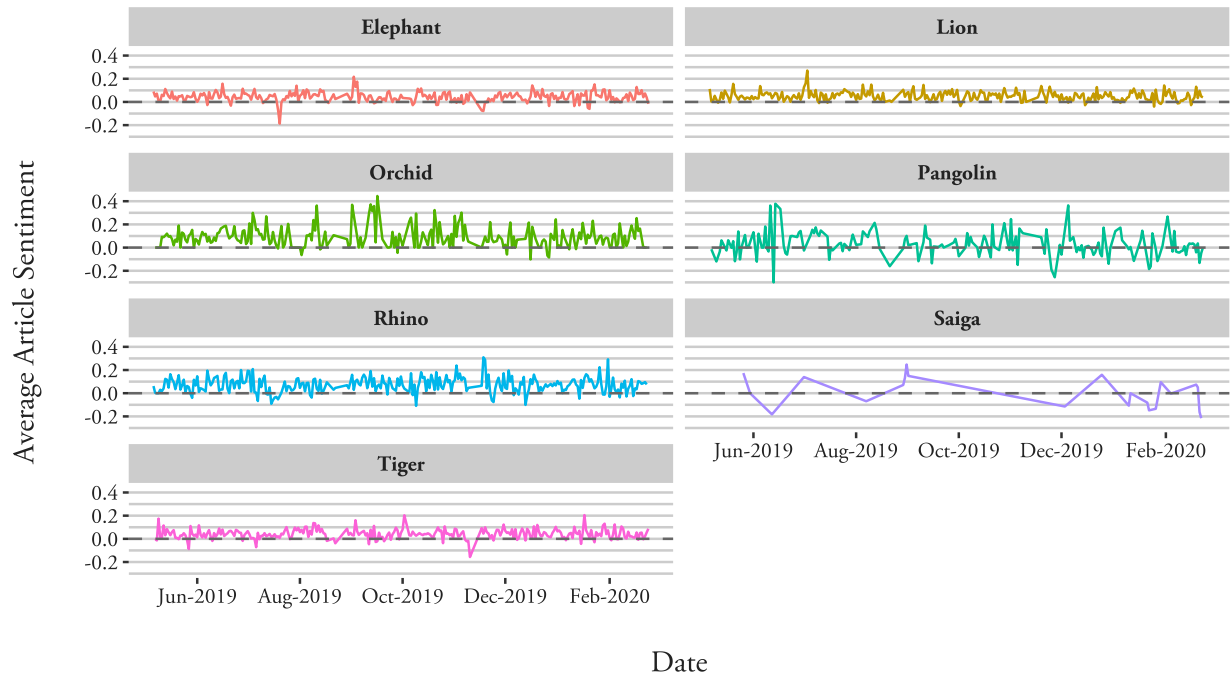

*Figure 2: Mean article sentiment per day for target taxa based on the Global Database of Events, Language, and Tone (GDELT; www.gdeltproject.org)*

The narrow range of taxa such as elephants, which are heavily involved in human wildlife conflict, suggests that reporting around those issues is not substantially captured in our dataset or that it is obscured by a dominance of media sources from non-range countries where these events do not take place. This is expected given the bias in the number of sources included in GDELT, which has a higher representation of sources from North America and Europe.

## Spatial Trends

Media coverage was globally distributed, with articles from 73 countries across all continents, which is not surprising given the wide and often non-overlapping ranges of our focus taxa. Elephants and tigers were the most widely covered with reporting in outlets of 59 and 54 countries respectively. On the other hand orchids and saiga received the least coverage, appearing in outlets of 25 and 10 countries respectively. These results are perhaps unsurprising, as elephants and tigers are very widely distributed. It does, however, corroborate the notion that plants tend to receive less coverage: orchids, being a large taxa with thousands of species, are by far the most widely distributed in this study but received comparably little coverage.

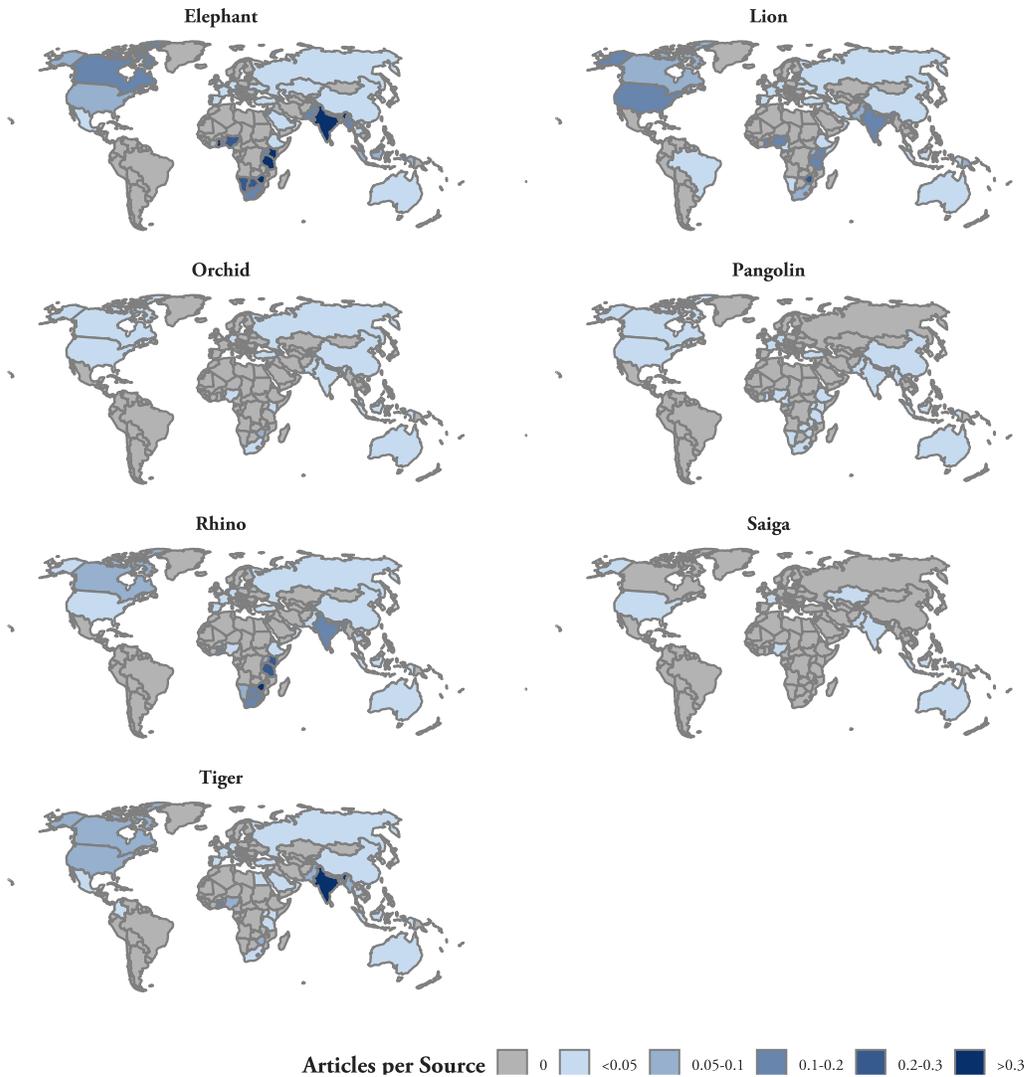

*Figure 3: Number of articles relevant per target taxa per news source on the Global Database of Events, Language, and Tone (GDELT; www.gdeltproject.org)*

Elephants, lions, rhinos, and pangolins received the most coverage in outlets from their range countries in Africa, with Zimbabwe featuring in the top three for all four groups. For tigers, the most coverage per outlet came from India, where they are native, but the second and third place were taken up surprisingly by Ghana and the United Arab Emirates, both non-range countries. For saiga, the country with the most coverage per outlet was Kyrgyzstan, where the species does not occur. Orchids, which are globally distributed, had most coverage in India, Jamaica, and Zimbabwe. These results support the notion that African megafauna have a high profile across their range countries, which is not surprising given how many of these species are not only involved in human wildlife conflict but are also key in the ecotourism industries of many of these countries, and therefore key for the national economy. India was the only country to be present in top ten of most coverage across all target taxa, with Zimbabwe being present in all taxa except saiga. This highlights the importance of wildlife in these countries, which face great challenges both in terms of economic development and wildlife management.

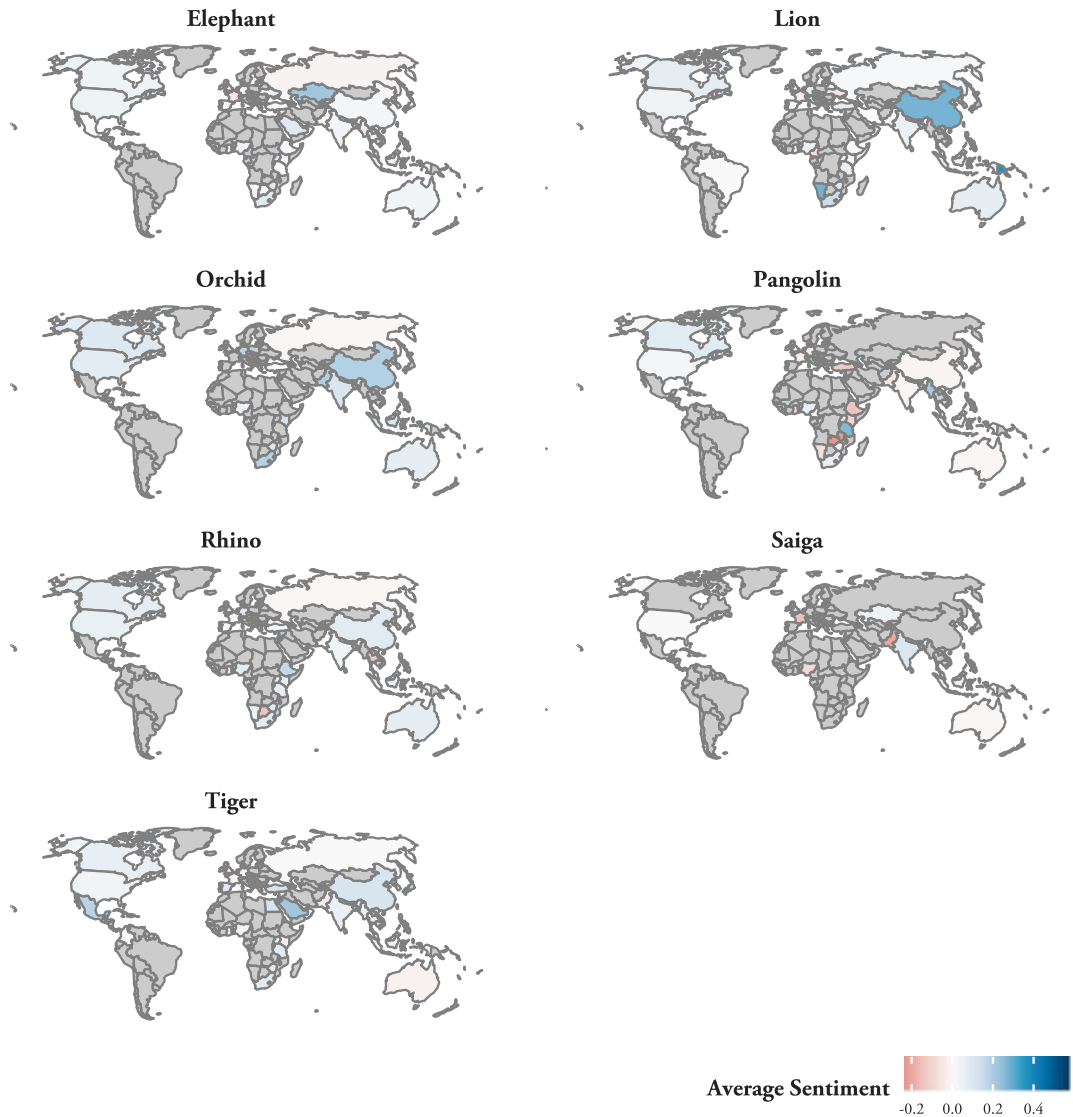

*Figure 4: Mean sentiment for each target taxa by country on the Global Database of Events, Language, and Tone (GDELT; www.gdeltproject.org)*

When considered in spatial terms, mean sentiment towards charismatic megafauna was most positive in non-range countries, with Jamaica surprisingly coming up repeatedly as a top three country for positive sentiment towards rhino, elephant, and tiger. The exception to this rule were pangolins and orchids, for which the countries with most positive sentiment were range states. For pangolins this is positive, suggesting that the positive attitudes towards the species can help drive greater conservation measures required by the taxon. For orchids this is less meaningful, as they have a global distribution and so would be expected to receive both the highest degree of positive and negative coverage in range countries.

Sentiment was most negative also in non-range countries, with France and Syria featuring twice among the top three countries with most negative mean sentiment across all taxa. Surprisingly, tigers received the most mean negative sentiment in non-range countries, with lions and elephants also having a majority of non-range states in their top three countries with the most negative sentiment score. As above, pangolins and orchids were the exception, with most countries in their top three for negative sentiment being range countries. For pangolins these countries did not match those where seizures often occur, for example Vietnam, and so this negative coverage does not appear to be a direct result of reporting on these events. As above, given the wide distribution of orchids, these results are hard to interpret.

## Why Societal Value Matters

Conservation is inherently an interdisciplinary science that includes subdisciplines such as conservation psychology, which acknowledges the role that human conventions and societal pressures play in allocating conservation efforts (Saunders 2003). Understanding the risk of extinction of a species is therefore more than understanding the ecology that underpins its life cycle and that of its habitat. The probability of a species going extinct is heavily influenced by the value that societies place on that species, including both market and non-market values (Cazabon-Mannette et al. 2017). There are multiple causal pathways through which societal forces can affect species at risk of extinction. Some forces are direct, such as when a species is persecuted, as in the case of unregulated hunting or illegal fishing (Gallic & Cox 2006) or conversely when people actively support a species' persistence, for example by providing food or other resources (Hejcmanová et al. 2013). Other pathways are indirect, like the provision of additional funding, from either government or private sources, or a change in policy.

Societal values vary enormously among taxa, which translates into disparities in funding and legal protection.

In fact, they vary to the extent that only mammals, birds and amphibians have had global conservation status assessments, despite these groups making up only a tiny fraction of the world's biodiversity. Many taxa, for example among invertebrates and fungi, have limited to no understanding of the scale of the threats they face, which is a key part of any meaningful large-scale effort to conserve them (Dahlberg et al. n.d.; Dahlberg & Mueller 2011; Cardoso et al. 2011). Consequently, efforts to conserve lesser appreciated species often resort to the umbrella species or flagship species concepts, which places visible and charismatic species as the direct recipient of attention and funds that will have trickle-down benefits for the species that are more difficult to generate appreciation for (Roberge & Angelstam 2004). Flagship species are often chosen to resonate with the public: those with large eyes, and that can easily be marketed as plush toys (Home et al. 2009). The challenge faced by conservation practitioners and organizations is to ensure representation of ecosystems and biomes by using these campaign species so that the benefits are broad. Ensuring that societal values are capitalized when designing conservation programs and allocating limited funds is a great challenge of conservation. Limited data are available to evaluate how well flagship or umbrella species concepts work to reverse declines of species under the umbrella, but it is generally perceived as one of the best available tools for gaining support for conservation actions.

## Future challenges

The availability of real time conservation human dimensions data through digital and social media is rapidly revolutionizing how conservation can be practiced. Mining these novel data sources for access to human experience and sentiments towards wildlife can greatly enhance the efficiency of conservation efforts by identifying hotspots for positivity and areas where campaigns are needed to improve science and natural history literacy. Ultimately, our study provides a window into how data on human attitudes towards wildlife have the potential to change how we understand the human dimensions of wildlife conservation.

We envision an upscaling of these tools for monitoring conservation psychology, yet there are no doubt challenges to the broader use of natural language processing for conservation science and practice. The first is that it would require the adaptation of the tools currently used by marketers to the context of biodiversity conservation, as most have been developed with business aims, highlighted by how, for example, sentiment analysis lexicons associate some animals (e.g. sharks, parasites) with negative sentiments (Lennox et al. 2020). The second is that there may be constrains to accessing some key information sources, particularly some of the larger social media platforms that have recently discontinued public application programming interfaces (APIs). New sources of data will continue to emerge, however, requiring additional methodological development and curation. Third, there will be biases in the data that are difficult to test. There is disparate access to the internet in different regions of the world and among socio-economic groups. Mitigating against these disparities would require a range of collaborations among the biological, social, computer, and statistical sciences to overcome biases and account for differences in representation.

Despite the challenges, there is a clear path forward for integrating digital and social media data in conservation plans to enhance efforts to track global trends and avenues for improving conservation outcomes worldwide. As more data are collected across time, the data can begin to reveal cause and effect relationships across time and space. Counterfactual-based causal inference methods such as the synthetic control technique [12], which uses longitudinal data to understand cause and effect relationships, could help to understand the impact of geopolitical events across the world on the relationships between society and wildlife. These methods can yield the data needed to test, for example, how political regime changes, war, and pandemics influence the focus on conservation and attitudes towards nature. This approach could be used in the medium term as an early warning system, for example in cases of emerging conflicts between people and wildlife as well as shifts in demand for wildlife products.

*The Digital Divide*

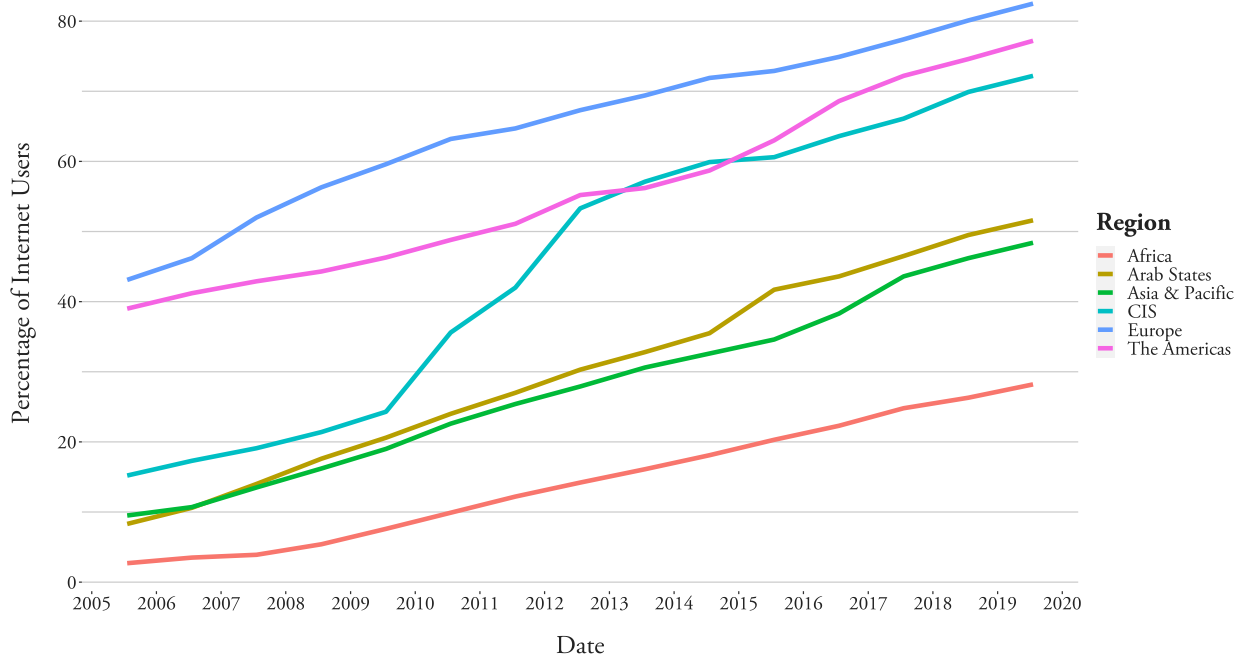

*Figure 5: Percentage internet penetration by region. Source: ITU World Telecommunication/ICT Indicators database.*

While the access to the internet has been growing steadily since the turn of the millennium, access to the internet is not equal across the globe, with higher income countries having nearly twice the internet penetration of lower income countries (**Error! Reference source not found.**). This reality is reflected in the sources upon which GDELT is based, as about two thirds of sources are from North America and Europe. This inequality is, however, likely to be reduced in the coming decade as higher income countries reach a ceiling in terms of internet penetration, and smartphone use continues to expand in lower income countries, in which 3G networks or above already cover 92% of the population (International Telecommunications Union 2019).

The digital divide is also visible at a smaller scale across socio-economic status, gender, and age in that poorer, female, and older people are less likely to use the internet (Chen & Wellman 2004). This bias impacts not only the information generated by online users, for example on social media, but also news sources: news in countries with low levels of internet penetration is more likely to rely on traditional print media for dissemination, making it largely unavailable to online study. These biases have important consequences for any analysis of global sentiment using online media as views towards different wildlife species are likely to vary across countries with, for example, citizens of western countries known to favour mammalian megafauna (Macdonald et al. 2015). This bias can be mitigated by weighting data from less representative sources, and approach that has been used extensively in other fields to try to counter response bias (Little & Vartivarian 2003). This approach has its own challenges, however, as calculating appropriate weightings relies on its own sets of data and assumptions to avoid replacing one set of biases with another.

### *Diversity of Data Sources*

Diversity of sources, and understanding the relative bias of these sources, is crucial to studies of online opinion. To support our gathered data with statistics on global use of the internet, we use the Alexa web traffic and ranking dataset, a well-established, global dataset of website popularity managed by Alexa Internet since 1996, and ultimately owned by Amazon since 1999. Historically, website popularity data was calculated from a user-installed browser toolbar that reported aggregate statistics about users' browsing behaviours directly; in recent years, Alexa has relied both on this traditional browser toolbar and a range of other internal measurement sources, including direct measurement code installed by website owners (Alexa n.d.).

To assess the relative impact of social media usage and linguistic diversity amongst websites, and thus the sources available to GDELT, we make use of the *cloudyr* R API for Amazon's Web Services (cloudyr project n.d.) to retrieve a ranked listing of the top 100 websites for each country globally. Whilst this data does not provide raw traffic volumes, the Alexa dataset uses a relative score to rank pages according to visits. The Alexa data also provides categorisation of the URLs, allowing us to identify news, social media, shopping, and other categories.

Using the raw list of URLs for each country, we can make use of the Google Translate API to identify the language presented by the landing page for each URL in the top 100. The combined dataset therefore reveals the nature of the most popular websites by user visits in each country globally, as well as their dominant language. With this in hand, we can identify the key languages used when accessing information online in each country, as well as the diversity of social media sites frequented by users.

Recent decades have seen the progressive replacement of offline news by online sources as traditional media sources, such as newspapers and television stations, move their content online, thus positioning the internet as the key news channel. At the same time, as an entirely new phenomenon alongside traditional news dissemination, we have seen the growing importance of social media as a source of information, with billions of users producing content (Nielsen & Schrøder 2014). These changes presents a tremendous data source for researchers trying to understand how attitudes and social norms vary across time and space, however accessing these different sources can be fraught with challenges both legal and technical.

Twitter is a common focus of studies of online opinion not only because the service is focused around opinion and discussion, but also significantly because it provides easy programmatic access to its data in contrast to many social media competitors. As such, whilst Twitter provides a useful base dataset for mapping public opinion, there is a need for means to access alternative services and understand the biases inherent in their user populations.

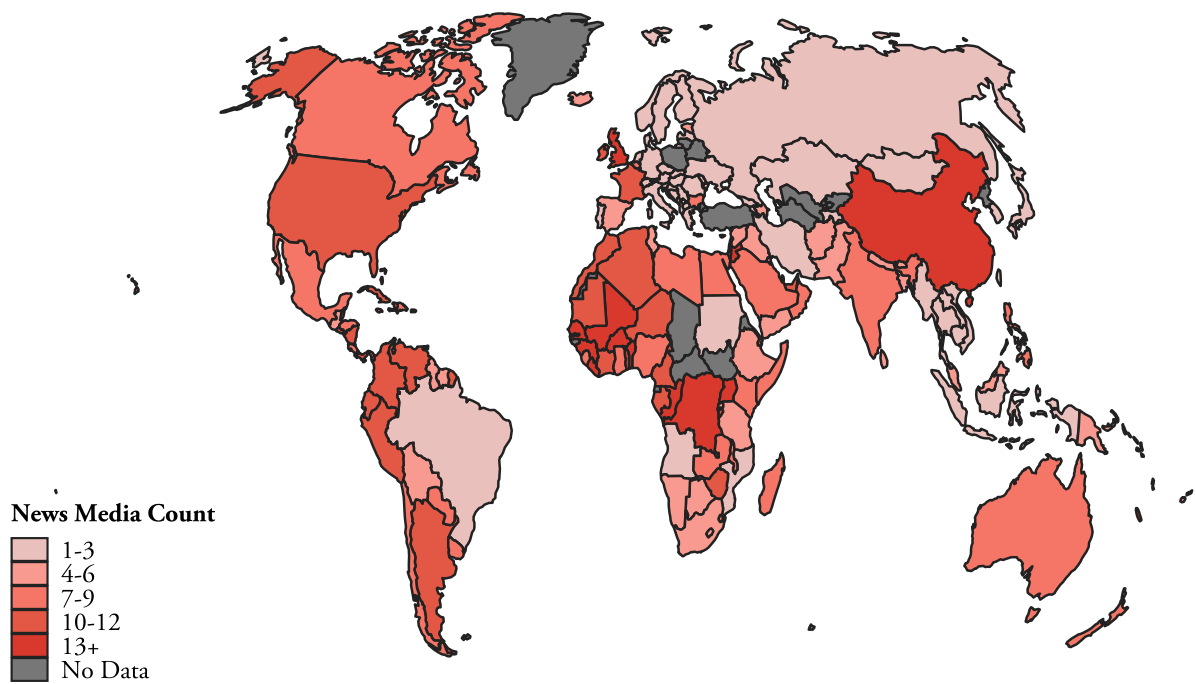

*Figure 6: Number of news media websites amongst top 100 websites per country.*

***Linguistic Diversity***

Sentiment analysis tools have been developed for a number of languages, however the large majority of widespread tools focus solely on English. Those tools and datasets that do focus on other languages are themselves single-language focused. This lack of a multi-language, or language agnostic, mechanism for analysing comparitive sentiment at a global scale is a major limitation to current techniques. As we demonstrate here, there are a huge diversity of languages being used online; to be able to gain a true global understanding we will need to analyse data across languages.

This problem is more than simple country-level representation: within and across countries different demographic groups make use of different online platforms in different languages, or express themselves in multiple languages within one platform, depending on sociolinguistic context. Even beyond the basic problem of discrimination introduced by internet penetration, overrepresenting higher-income demographic groups, this bias *within* the available results should be cause for careful consideration.

It should be noted that whilst these limitations of online data for broad-scale sentiment analysis are significant, they are in no way absent from alternative offline methods that also struggle with bias and representation without the potential for ongoing analysis at the global scale offered by online methods.

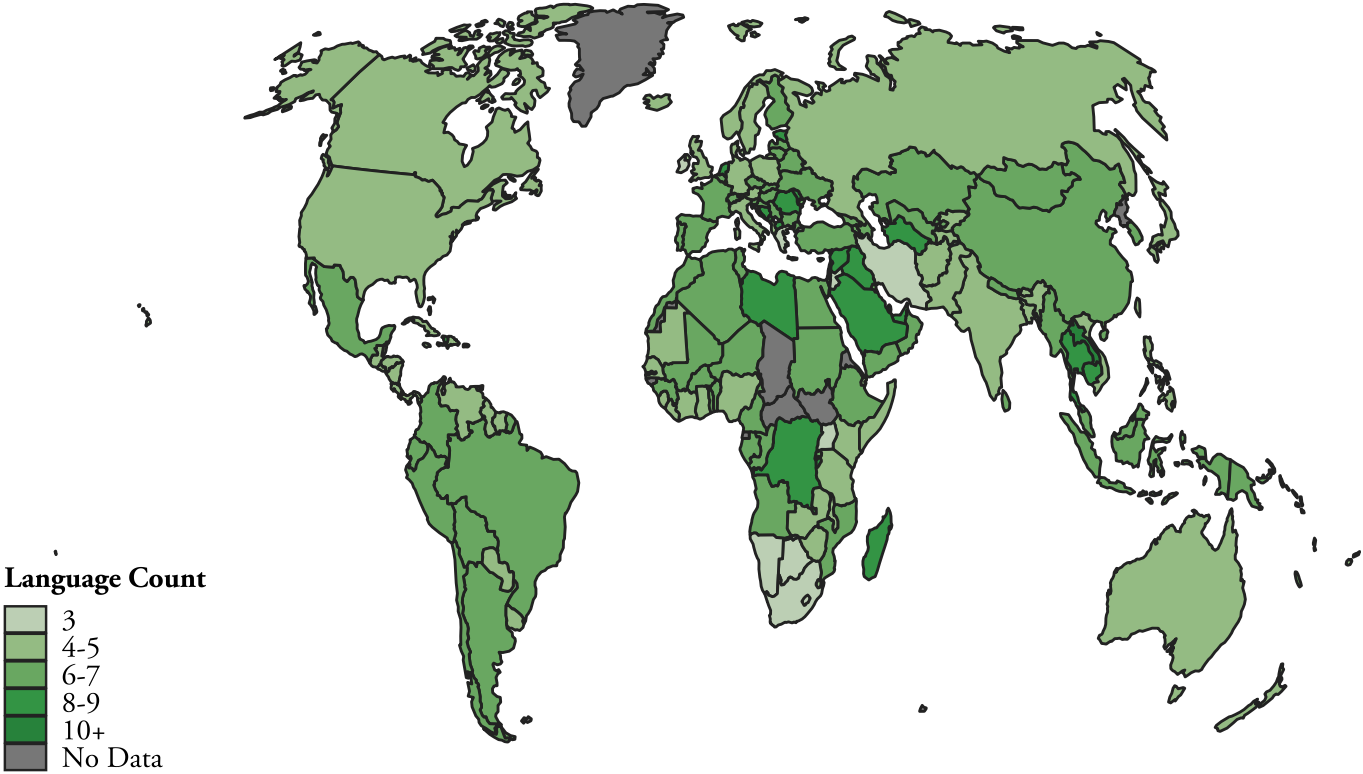

*Figure 7: Number of languages amongst top 100 websites per country.*

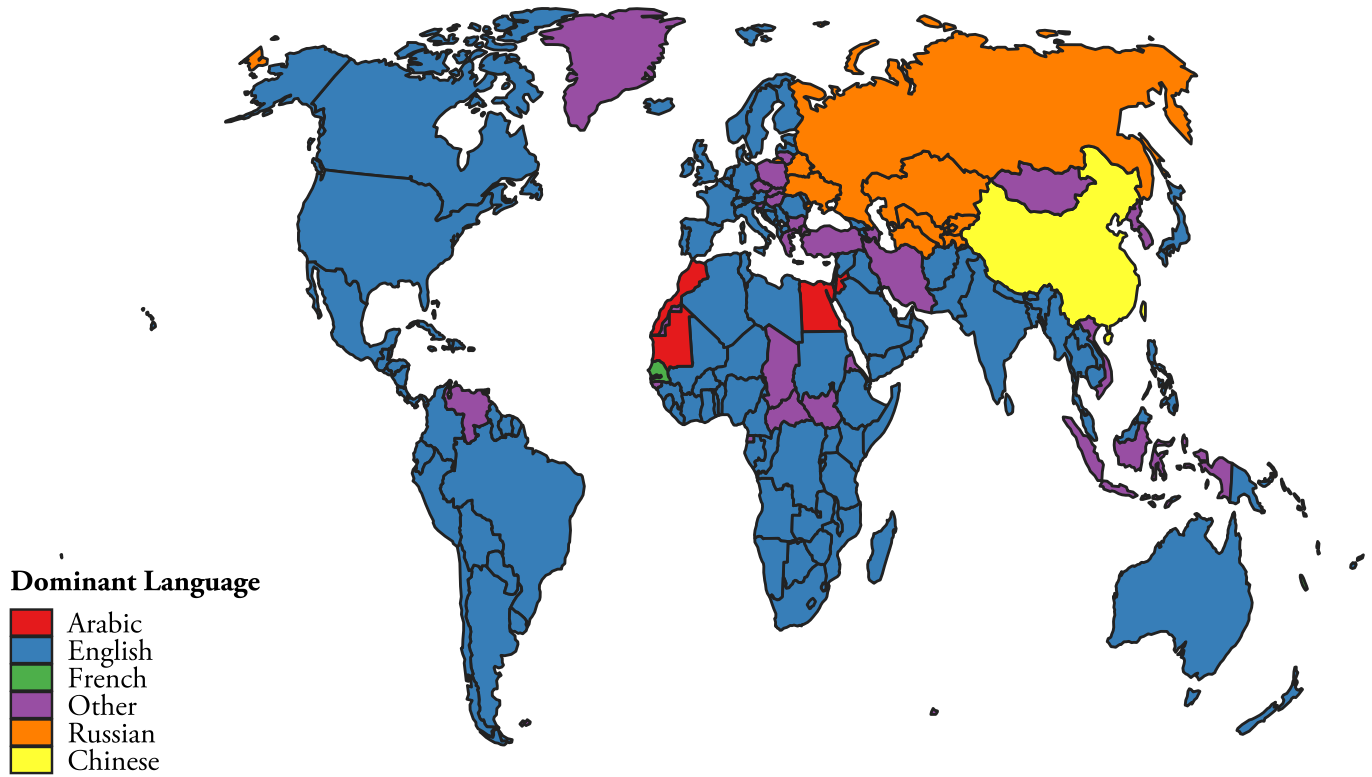

*Figure 8: Most common language amongst top 100 websites per country.*

## Conclusion

The global adoption of the internet has brought with it tremendous potential for the monitoring of human attitudes and social norms regarding wildlife in a way that has thus far remained largely unexplored. While there remain many constraints, not only in the data generation process but also in the analysis and interpretation of these large datasets, online data can provide insights at a temporal and spatial scale far beyond the scope of alternative approaches. The known shortcomings of online approaches, such as geographic and demographic bias in internet access, are likely to be mitigated only slowly and are largely beyond the control of researchers.

The method developed here, making use of open-access large-scale data sources, demonstrates the use of internet-based data as a means to maintain a global view of wildlife sentiment. Free and open data makes this approach widely accessible to researchers, but comes with caveats regarding diversity and representation. In addition, due to the limitations of search and the scale of the datasets in question, we demonstrate that the use of machine learning techniques to automate the cleaning and selection of data is necessary for meaningful results.

The development of improved sentiment analysis tools specifically for use in the context of wildlife conservation present the opportunity to gain more accurate understanding of human perceptions of wildlife and biodiversity. The application of other techniques drawn from machine learning and natural language processing, such as topic modelling and time series anomaly detection, will allow us to gain a more robust understanding of how sentiment should be interpreted in the context of conservation, and how and where coverage shifts in relation to both offline and online events. The combination of these approaches within the social sciences, drawing on computer and data science, will surely prove to be fertile ground for future work in the emerging field of conservation culturomics.

## Appendix: Social Media Landscape

In addition to news media, social media interactions are similarly dominated by a small number of platforms. The availability of data scraping tools to access these platforms, and the discussions that take place on them, significantly skews the representativeness of social media data analysis approaches.

*Figure 9: Count of social media websites amongst the top 100 websites per country.*

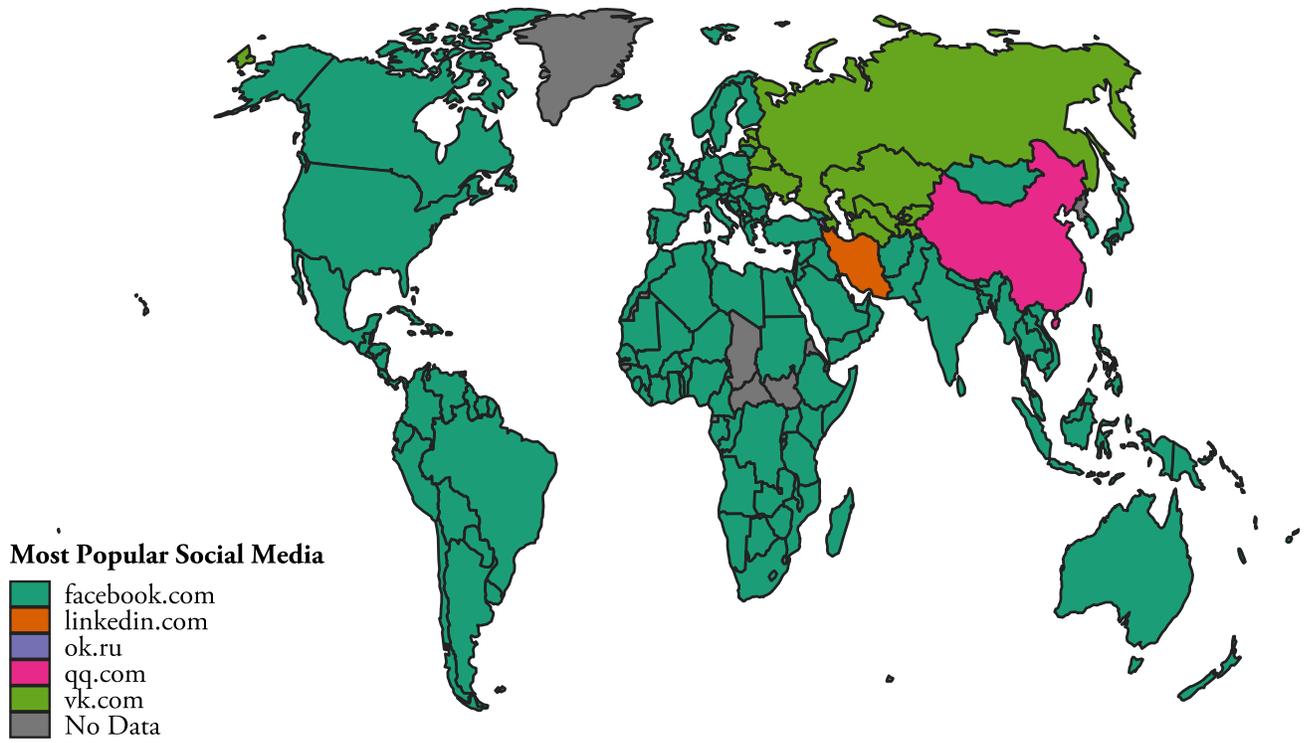

*Figure 10: Most popular social media site by country.*